\documentclass[aps,prc,superscriptaddress,reprint]{revtex4-1}
\usepackage{tabularx}
\usepackage{graphicx} 
\usepackage[strings]{underscore}
\usepackage{CJKutf8}
\usepackage{epstopdf}
\usepackage{dcolumn} 
\usepackage{bm}
\usepackage{hyperref}
\usepackage{array}
\usepackage{setspace}
\usepackage{booktabs}
\usepackage{amsmath}
\usepackage{mathtools}
\usepackage[section]{placeins}
\hypersetup{
    colorlinks=true,
    linkcolor=blue,
    filecolor=gray,
    urlcolor=blue,
    citecolor=blue,
}
\begin{document}
\begin{CJK}{UTF8}{gbsn}

\title{Exploring the potential of synthesizing unknown superheavy isotopes via cold-fusion reactions based on the dinuclear system model}
\thanks{This work is supported by National Science Foundation of China (NSFC) (Grants No. 12105241,12175072).}%

\author{Hao Wu(吴浩)}
\affiliation{School of Physical Science and Technology, Yangzhou University, Yangzhou 225009, China}

\author{Peng-Hui Chen(陈鹏辉)}
\email{Corresponding author: chenpenghui@yzu.edu.cn}
\affiliation{School of Physical Science and Technology, Yangzhou University, Yangzhou 225009, China}

\author{Fei Niu(牛菲)}
\affiliation{Henan Police College, Zhengzhou 450046, China}

\author{Zu-Xing Yang(杨祖星)}
\affiliation{RIKEN Nishina Center, Wako, Saitama 351-0198, Japan}

\author{Xiang-Hua Zeng(曾祥华)}
\affiliation{School of Physical Science and Technology, Yangzhou University, Yangzhou 225009, China}
\affiliation{College of Electrical, Power and Energy Engineering, Yangzhou University, Yangzhou 225009, China }

\author{Zhao-Qing Feng(冯兆庆)}
\email{Corresponding author: fengzhq@scut.edu.cn}
\affiliation{School of Physics and Optoelectronics, South China University of Technology, Guangzhou 510641, China}

\date{\today}
\begin{abstract}
To assess the potential of cold-fusion for synthesizing superheavy nuclei (SHN) with proton numbers 104-113, we systematically calculated 145 naturally occurring projectile-target combinations within the dinuclear system (DNS) model. Reactions predominantly show maximum cross sections in the 1n to 2n channels, peaking near the Coulomb barrier with a sum of barrier and $Q$-value within 30 MeV. The maximum cross section occurs below the Bass barrier, suggesting either the Bass model's limitation or significant deformation reducing the effective Coulomb barrier. Our calculations align well with experimental data, revealing that more neutron-rich projectiles slightly enhance fusion, though the effect is minor. For fixed targets (Pb or Bi), evaporation residue cross sections decrease linearly with increasing projectile proton number, attributed to reduced fusion probability and lower fission barriers in heavier SHN. The contact potential $V_{\rm in}$ shows a linear trend with the product of projectile-target proton numbers, with neutron-rich systems exhibiting lower $V_{\rm in}$. Some reactions with $V_{\rm in} < V_{\rm S}$ may involve nucleon transfer before capture. Based on the DNS model, we identify optimal combinations and collision energies for synthesizing SHN with significant cross sections. Collectively, our findings indicate that cold fusion is a promising avenue for creating proton-rich SHN around the drip line in the $Z$=104-113 region, offering distinct advantages over alternative mechanisms.
\begin{description}
\item[PACS number(s)]
25.70.Jj, 24.10.-i, 25.60.Pj
\end{description}
\end{abstract}

\maketitle

\section{Introduction}

The synthesis of superheavy nuclei (SHN) represents one of the most challenging and intriguing frontiers in nuclear physics\cite{Thoennessen16}. 
The concept of superheavy elements was first theorized in the 1960s when researchers like Sobiczewski, Gareev, and Kalinkin proposed that there might be a ``superheavy stable island" around the atomic number $Z$=114 and neutron number $N$=184, based on the nuclear shell model\cite{SOBICZEWSKI1966500}.
Among the various experimental approaches, cold fusion reactions have emerged as a promising pathway to extend the chart of nuclides towards the superheavy regime. Cold fusion is characterized by the merging of two heavy nuclei at relatively low energies, typically below the Coulomb barrier, which circumvents the traditional high-energy fusion routes. 
In 1994, GSI (Gesellschaft f\"{u}r Schwerionenforschung) in Germany pioneered the synthesis of superheavy elements 110, 111, and 112 via cold fusion of heavy ions, ushering in a new epoch in the study of heaviest elements\cite{RevModPhys.72.733}. Subsequently, laboratories globally advanced cold fusion methodologies. Notably, the Russian Dubna/FLNR and Japan's RIKEN have contributed significantly to this domain.
In 2004, RIKEN's Japanese team synthesized element 113 (Nihonium) via cold fusion of $^{209}$Bi and $^{70}$Zn, marking a significant expansion of the periodic table\cite{2004}. The experiment observed a mere three decay events, with a cross section of 0.02 pb, highlighting the potential for superheavy element creation.
This method offers a unique opportunity to explore the limits of nuclear stability and the underlying structure of the atomic nucleus.
The quest to synthesize SHN through cold fusion is not only driven by the pursuit of scientific knowledge but also by the potential applications of these exotic species in various fields, including nuclear medicine, energy production, and environmental remediation\cite{TianJunLong2007}. As our understanding of the underlying physics deepens and experimental techniques advance, the prospect of discovering new superheavy elements and unraveling the mysteries of the superheavy island of stability becomes increasingly tangible. 

Despite the low probabilities associated with cold fusion, the potential to create new elements and isotopes that are far from the valley of stability has spurred extensive research in this field. 
Despite the progress, synthesizing superheavy elements remains a challenging task due to the extremely low cross sections for fusion and the short half-lives of the produced nuclei. However, each successful synthesis not only expands our knowledge of the periodic table but also provides valuable insights into the fundamental properties of atomic nuclei.
The competition in the field of synthesizing new SHEs with the atomic number 119-120 is fierce. China IMP, Russia JINR, and Japan RIKEN have all built new experimental devices and launched synthesis plans\cite{Gan2022}.
However, the experimental synthesis of SHN is a challenging, time-consuming, and costly process, often yielding extremely small cross sections. Therefore, reliable theoretical studies that can guide experiments are crucial. In this paper, the dinuclear system (DNS) model has been instrumental in providing a theoretical framework to understand the complex dynamics of cold fusion reactions. It accounts for the coupled motion of two interacting nuclei, the energy dissipation mechanisms, and the subsequent decay processes that lead to the formation of compound nuclei.
These models helped predict the optimal combinations of projectile and target nuclei, as well as the necessary incident energies for synthesizing unknown SHN \cite{FENG200650,PhysRevC.91.011603,PhysRevC.89.024615,epja20gga,07fengcpl}.

Compared to hot-fusion reactions, cold-fusion reactions for the synthesis of SHN are characterized by lower excitation energies of the compound nucleus and maximum cross sections in the 1-2 neutron emission channels, with corresponding collision energies below the Coulomb barrier. The advantage of using cold fusion reactions for synthesizing SHN with atomic numbers Z=104-113 is the presence of larger synthesis cross sections. This is demonstrated by the fact that, although the fusion probabilities are low, the survival probabilities through the 1-2 neutron emission pathways are relatively high. However, since the maximum cross section corresponds to collision energies below the Coulomb barrier, calculating the fusion probability under the barrier is crucial for accurately reproducing and predicting the synthesis cross sections of SHN. Yet, due to the complex dynamical characteristics of sub-barrier fusion, it remains inexact to this day.
In this study, we employ the DNS model to calculate the cold-reactions, the selections of $^{48}$Ca, $^{50}$Ti, $^{52,54}$Cr, $^{58}$Fe, $^{59}$Co, $^{62,64}$Ni, $^{70}$Zn, $^{51}$V, $^{55}$Mn as projectiles, and $^{204,206,207,208}$ Pb, $^{209}$Bi as the targets, benchmark our model against the available cold-fusion experimental results to find the optimal contact potential energies for improvement.
Based on the DNS model and the optimal contact potential energies, we make predictions for synthesis cross sections of the unknown SHN via cold fusion reactions, systematacially. 

\section{Model description}\label{sec2}

Volkov et al.'s DNS concept pioneered the understanding of deep-inelastic heavy-ion collisions \cite{VOLKOV197893}. Adamian's work subsequently integrated quasi-fission into the fusion paradigm \cite{ADAMIAN1997361,ADAMIAN1998409}. The model was further refined in Lanzhou, incorporating the interplay of kinetic energy and impact parameters with nucleon rearrangement \cite{PhysRevC.76.044606}. It outlines a pathway involving capture, fusion, and survival of excited compound nuclei, pivotal for SHN synthesis \cite{FENG200650,22nstma}.
Based on the DNS model, the evaporation residual cross section of SHN can be written as
\begin{eqnarray}
\sigma _{\rm ER}\left ( E_{\rm c.m.} \right ) =\frac{\pi \hbar ^2}{2\mu E_{\rm c.m.}} \sum_{J=0}^{J_{\rm max}}(2J+1)T(E_{\rm c.m.},J)\nonumber \\ \times P_{\rm CN}(E_{\rm c.m.},J)W_{\rm sur}(E_{\rm c.m.},J). \end{eqnarray}
Here, $T$ is the probability of the collision system passing through the Coulomb barrier \cite{FENG200650}. The $P_{\rm CN}$ is the fusion probability \cite{PhysRevC.80.057601,PhysRevC.76.044606}. The $W_{\rm sur}$ is the probability of survival. Here, we take the maximal angular momentum as $J_{\rm max}$ = 100 $\hbar$, because the fission barriers of compound nuclei and the quasi-fission barrier of the composite system fade away at high spin \cite{J.Mod.Phys.E5191(1996)}.

\subsection{ Capture probability}

The capture cross section is evaluated by
\begin{eqnarray}
\sigma _{\rm cap}(E_{\rm c.m.})=\frac{\pi \hbar ^2}{2\mu E_{\rm c.m.}}\sum_{J}^{}(2J+1)T(E_{\rm c.m.},J).
\end{eqnarray}
Where the penetration probability $T(E_{\rm c.m.},J)$ is evaluated by the Hill-Wheeler formula \cite{PhysRev.89.1102} and the barrier distribution method.

\begin{eqnarray}\label{hwl}
&T(E_{\mathrm{c.m.}},J)=\displaystyle\int \frac{f(B) \mathrm{d}B}{1+\exp\Big\{-\frac{2\pi (E_{\mathrm{c.m.}}-B-E^{\rm rot})}{\hbar\omega(J)}\Big\}}.
\end{eqnarray}

Here $\hbar \omega (J)$ is the width of the parabolic barrier at $R_{\rm B}(J)$. The normalization constant is $\int f(B)\rm{d}B=1$. The barrier distribution function is the asymmetric Gaussian form  \cite{FENG200650,PhysRevC.65.014607}
\begin{eqnarray}\label{asg}
f(B)=
\left\{\begin{matrix}
\frac{1}{N}\exp [-(\frac{B-B_{\rm m}}{\Delta_{1} } )]\ \ B<B_{\rm m},\\
  & \\\frac{1}{N}\exp [-(\frac{B-B_{\rm m}}{\Delta_{2} } )]\ \ B>B_{\rm m},
  &
\end{matrix}\right.
\end{eqnarray}
Here $\bigtriangleup _2=\left ( B_0-B_{\rm S} \right )/2$, $\bigtriangleup_1=\bigtriangleup_2-2$ MeV, $B_{\rm m}=(B_0+B_{\rm s})/2$. \( B_0 \) and \( B_{\rm s} \) represent Coulomb barriers at distinct stages: \( B_0 \) arises at the contact point in a side-by-side collisions, and \( B_{\rm s} \) is encountered at the potential well's nadir where variations in quadrupole deformation are considered. 
The interaction potential of two colliding partners is written as 
\begin{eqnarray}\label{vcn}
 V_{\rm CN}(\{\alpha\}) = && V_{\rm C}(\{\alpha\}) + 
V_{\rm N}(\{\alpha\}) \nonumber \\ 
&& +\frac{1}{2} C_{1}(\beta_{1}-\beta_{1}^{0} )^{2}+\frac{1}{2} C_{2}(\beta_{2}-\beta_{2}^{0} )^{2}.   
\end{eqnarray}
The $\{\alpha\}$ stands for $\{r,\beta_{1},\beta_{2},\theta _{1},\theta_{2}\}$.
The subscript 1 and 2 stand for the projectile and the target, respectively. The $R=R_1+R_2+s$ and s are the distance between the center and surface of the projectile-target. The $R_1$, $R_2$ are the radii of the projectile and target, respectively. The $\beta_{1(2)}^0$ are the static quadrupole deformation. The $\beta_{1(2)}$ are the adjustable quadrupole deformation\cite{22nstma,MOLLER20161}. 
The nucleus-nucleus potential is calculated by the double-folding method  \cite{PhysRevC.80.057601,PhysRevC.76.044606, J.Mod.Phys.E5191(1996)}.
The Coulomb potential is evaluated by Wong's formula \cite{PhysRevLett.31.766}.

\subsection{Fusion probability}

The time evolution of the mass distribution probability is described by the master equation. For a mass distribution function $P(Z_1,N_1,E_1,t)$ with a nucleon number of $A_1$ and an internal excitation energy of $E_1$, the following equation is satisfied\cite{PhysRevC.76.044606,FENG201082,FENG200933}:
\begin{eqnarray}
&&\frac{d P(Z_1,N_1,E_1,t)}{d t} = \nonumber \\ && \sum \limits_{Z'_1}W_{Z_1,N_1;Z_1,N'_1}(t) [d_{Z_1,N_1}P(Z'_1,N_1,E'_1,t) \nonumber \\ && - d_{Z'_1,N_1}P(Z_1,N_1,E_1,t)] + \nonumber \\ &&
 \sum \limits_{N'_1}W_{Z_1,N_1;Z_1,N'_1}(t)[d_{Z_1,N_1}P(Z'_1,N_1,E'_1,t) \nonumber \\ && - d_{Z_1,N'_1}P(Z_1,N_1,E_1,t)] - \nonumber \\
 &&[\Lambda ^{\rm qf}_{A_1,E_1,t}(\Theta) + \Lambda^{\rm fis}_{A_1,E_1,t}(\Theta)]P(Z_1,N_1,E_1,t).
\end{eqnarray}
Here the $W_{\rm Z_1,N_1,Z'_1,N_1}$ is the mean transition probability from the channel ($Z_1,N_1,E_1$) to ($Z'_1,N_1,E'_1$), and $d_{\rm Z_1,N_1}$ denotes the microscopic dimension corresponding to the macroscopic state ($Z_1,N_1,E_1$). The sum is taking all possible proton and neutron numbers that fragment $Z'_1$, $N'_1$ may take, but only one nucleon transfer is considered in the model with the relation $Z'_1$ = $Z_1$ $\pm$ 1, and $N'_1$ = $N_1$ $\pm$ 1. The excitation energy $E_1$ is the local excitation energy $\varepsilon^*_1$ with respect to fragment $A_1$, which is determined by the dissipation energy from the relative motion and PES of DNS \cite{PhysRevC.27.590}. The sticking time is described by the parametrization method of classical deflection function \cite{LI1981107}. 
The evolution of the DNS along the variable $R$ leads to the quasi-fission of the DNS. The decay probability we evaluate by the one-dimensional Kramers equation \cite{PhysRevC.68.034601,PhysRevC.27.2063}.

The potential energy surface (PES) of the DNS is given by
\begin{eqnarray}
&&U_{\rm dr}(\{\alpha\})=B_1+B_2-B_{\rm CN}+V_{\rm CN}(\alpha)+V_{\rm rot}^{\rm CN} 
\end{eqnarray}
Here, the set $\{\alpha\}$ corresponds to the same definition as presented in Eq. (\ref{vcn}). $B_{\rm i}$ ($\rm i$ = 1, 2) and $B_{\rm CN}$ are the negative binding energies of the fragment $A_{\rm i}$ and the compound nucleus A, respectively, in which the shell and the pairing corrections are included reasonably; $V_{\rm rot}^{\rm CN}$ is the rotation energy of the compound nucleus; the $\beta_{\rm i}$ represent quadrupole deformations of the two fragments; the $\theta_{\rm i}$ denote the angles between the collision orientations and the symmetry axes of deformed nuclei. The interaction potential between fragment 1$(Z_1, N_1)$ and 2$(Z_2,N_2)$ includes the nuclear, Coulomb, and centrifugal parts. 

During the reaction, as the composite system evolves, fragment probabilities diffuse across the PES, yielding a distribution characterized by $P(Z,N,E_1,t)$. Successful fusion occurs when all fragments surpass the Businaro-Gallone (BG) point, culminating in the formation of compound nuclei.
The inner fusion barrier, a critical obstacle in the fusion process, is delineated by the differential in the driving potential from the initial approach to the BG point. Therefore, Fusion probability is written as
\begin{eqnarray}
P_{\rm CN}(E_{\rm c.m.},V_{\rm in})=\sum _{Z=1}^{Z _{\rm BG}}\sum_{N=1}^{N_{\rm BG}} P(Z,N,E',V_{\rm in}).
\end{eqnarray}
Here $Z _{\rm BG}$ and $N_{\rm BG}$ are the proton number and neutron number of the BG point.
The $B$ is the potential energy at the initial contact point.

\subsection{Survival probability}

The compound nuclei formed by all the nucleons transfer from projectile nuclei to target nuclei with certain excitation energies. The excited compound nuclei were extremely unstable which would be de-excited by evaporating $\gamma$-rays, neutrons, protons, $\alpha$ $etc.$) against fission. The survival probability of the channels x-th neutron, y-th proton and z-alpha. Development of a statistical evaporation model based on Weisskopf's evaporation theory. \cite{Chen2016,FENG201082,FENG200933,Xin2021}
\begin{eqnarray}
&&W_{\rm sur}(E_{\rm CN}^*,x,y,z,J)=P(E_{\rm CN}^*,x,y,z,J) \times \nonumber\\&&  \prod_{i=1}^{x}\frac{\Gamma _n(E_i^*,J)}{\Gamma _{\rm tot}(E_i^*,J)} \prod_{j=1}^{y}\frac{\Gamma _p(E_j^*,J)}{\Gamma_{\rm tot}(E_i^*,J)} \prod_{k=1}^{z}\frac{\Gamma _\alpha (E_k^*,J)}{\Gamma _{\rm tot}(E_k^*,J)}, 
\end{eqnarray}
where the $E_{\rm CN}^*$ and $J$ were the excitation energy and the spin of the excited nucleus, respectively. The total width $\Gamma_{\rm tot}$ was the sum of partial widths of particle evaporation, $\gamma$-rays, and fission. The excitation energy $E_S^*$ before evaporating the $s$-th particles was evaluated by
\begin{eqnarray}
E_{s+1}^*=E_s^* - B _i ^n - B _j ^p - B_k ^\alpha - 2T_s
\end{eqnarray}
with the initial condition $E\rm_i^*$=$E_{\rm CN}^*$ and $s$=$i$+$j$+$k$. The $B_{\rm i}^n$, $B_{\rm j}^p$, $B_{\rm k}^\alpha$ are the separation energy of the $i$-th neutron, $j$-th proton, $k$-th alpha, respectively. The nuclear temperature $T_i$ was defined by $E_{\rm i}^*=\alpha T_{\rm i}^2-T_{\rm i}$ with the level density parameter $a$. The decay width of the $\gamma$-rays and the particle decay were evaluated with a similar method in Ref.  \cite{Chen2016}.
The $P(E_{\rm CN}^*, J)$ is the realization probability of evaporation channels.  
The $\Gamma _n(E_i^*,J)$, $\Gamma _p(E_i^*,J)$ and $\Gamma _{\alpha}(E_i^*,J)$ are the decay widths of particles n, p, $\alpha$, which are evaluated by the Weisskopf evaporation theory \cite{PhysRevC.68.014616}. The fission width $\Gamma_f(E^*,J)$ was calculated by the Bohr-Wheeler formula \cite{PhysRevC.80.057601,artza05,PhysRevC.76.044606}.

In our calculation, the fission barrier has a microscopic part and the macroscopic part which is written as
\begin{eqnarray}
B_{\rm f}(E^*,J)=B_{\rm f}^{\rm LD}+B_{\rm f}^{\rm M}(E^*=0,J)\rm exp(-E^*/E_{\rm D})
\end{eqnarray}
where the macroscopic part was derived from the liquid-drop model.
Microcosmic shell correction energy was taken from  \cite{MOLLER20161}. Shell-damping energy was taken as $E_{\rm D}=5.48A^{1/3}/(1+1.3A^{-1/3})$ MeV.

\section{Results and discussion}\label{sec3}

To harness the cold-fusion approach for synthesizing SHN with atomic numbers ranging from 104 to 113, we have endeavored to enhance our DNS model using outcomes from cold-fusion experiments. Extensive calculations have been conducted to forecast the production cross sections of SHN predicated on the refined DNS model.
In this study, the DNS model is employed, distinguished by 
Simplicity: it simplifies the complex interactions within the nucleus by treating the nucleus as two interacting entities and reducing the need to calculate all nucleon-nucleon interactions, which can be easier to model and simulate, thus saving computational resources; 
Flexibility: it allows for the incorporation of various nuclear forces and correlations, making it adaptable to different types of nuclei and reactions; it offers an intuitive understanding of nuclear dynamics; it can self-consistently consider all kinds of physical quantities to realize specific research contents directly.
Predictive power: it is well-suited for the study of SHN where other models may fail to accurately predict stability and behavior \cite{Chen2023,Chen20232,niu2020,21nstfq,Yu2018,PhysRevC.93.044615,PhysRevC.93.064610,PhysRevC.89.024615,PhysRevC.90.014612,PhysRevC.100.011601,PhysRevC.96.024610,PhysRevC.95.034323,PhysRevC.92.034612,2018chen}.

\begin{figure}[htb]
\includegraphics[width=0.999\linewidth]{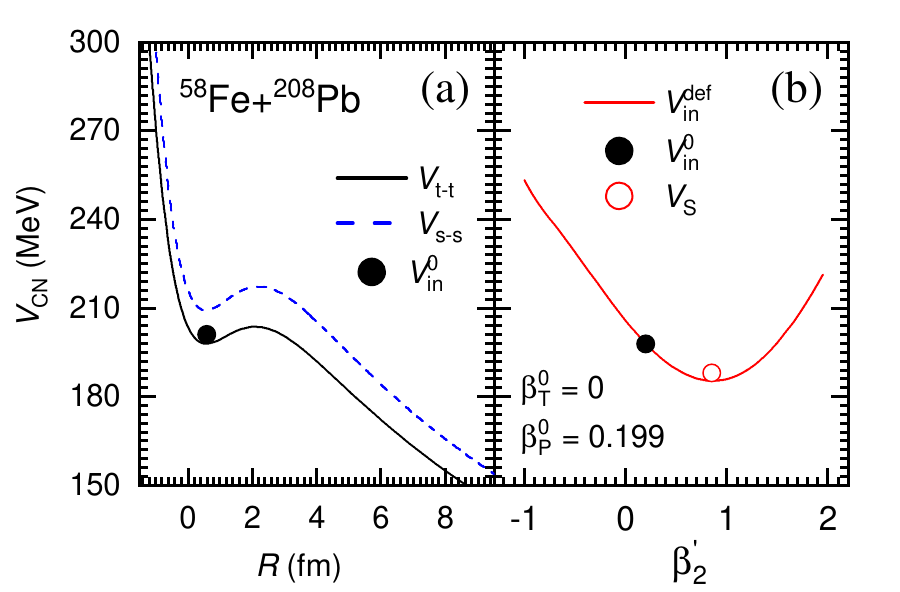}
\caption{\label{fig1} The interaction potential of $^{56}$Fe+$^{208}$Pb along the distance between the surface of projectile-target nuclei is listed in panel (a), evaluated by Eq. (\ref{vcn}). The solid black and blue dash lines stand for collision configurations of tip-tip and side-side, corresponding to $V_{\rm t-t}$ and $V_{\rm s-s}$. The black solid dot is the potential pocket bottom, represented by $V^0_{\rm in}$. Panel (b) shows the interaction potential varies by quadrupole deformation. The black open circle represents the potential energy at the pocket bottom with the initial shape of $\beta^0_{\rm P}$ and $\beta^0_{\rm T}$.
The red open circle is the minimum potential energy among the shape-dependent interaction potentials, marked by $V_{\rm S}$.
}
\end{figure}

Near-barrier heavy-ion collisions can initiate capture-fusion-evaporation reactions, which are inherently dynamic. During capture, the relative kinetic energy propels the collision partners to surmount the Coulomb barrier, forming a composite system evaluated by Eq. (\ref{hwl}). Damping collisions may induce dynamic deformation in the partners, and due to the random orientations, the Coulomb barriers exhibit an asymmetry-Gaussian-like distribution, calculated by Eq. (\ref{asg}). Nucleon transfer within the composite system occurs at the potential pocket's base, depicted in Fig. \ref{fig1}(a) and marked by $V^0_{\rm in}$, which is lower than the barrier energy.
The collision's orientation and dynamic deformation introduce variability in the contact point's potential energy. Panel (b) illustrates the interaction and deformation energies of the composite system. The solid circle represents the potential energy with the initial quadrupole deformation, equivalent to the solid circle in panel (a). The red open circle indicates the minimum potential energy observed during shape evolution.
As a crucial input in the DNS model, the contact point's potential energy dictates the kinetic energy dissipation within the composite system, directly influencing nucleon transfer. However, the contact potential energy's uncertainty spans from $V^{\rm W}_{\rm in}$ to $V^{\rm S}_{\rm in}$, representing energies at the side-side collision pocket's bottom and dynamic deformations, respectively. Previously, we addressed this uncertainty by employing a Gaussian-like distribution for the contact potential energies. In this study, we aim to identify the most effective $V_{\rm in}$ within this distribution range.

\begin{figure}[htb]
\includegraphics[width=1.\linewidth]{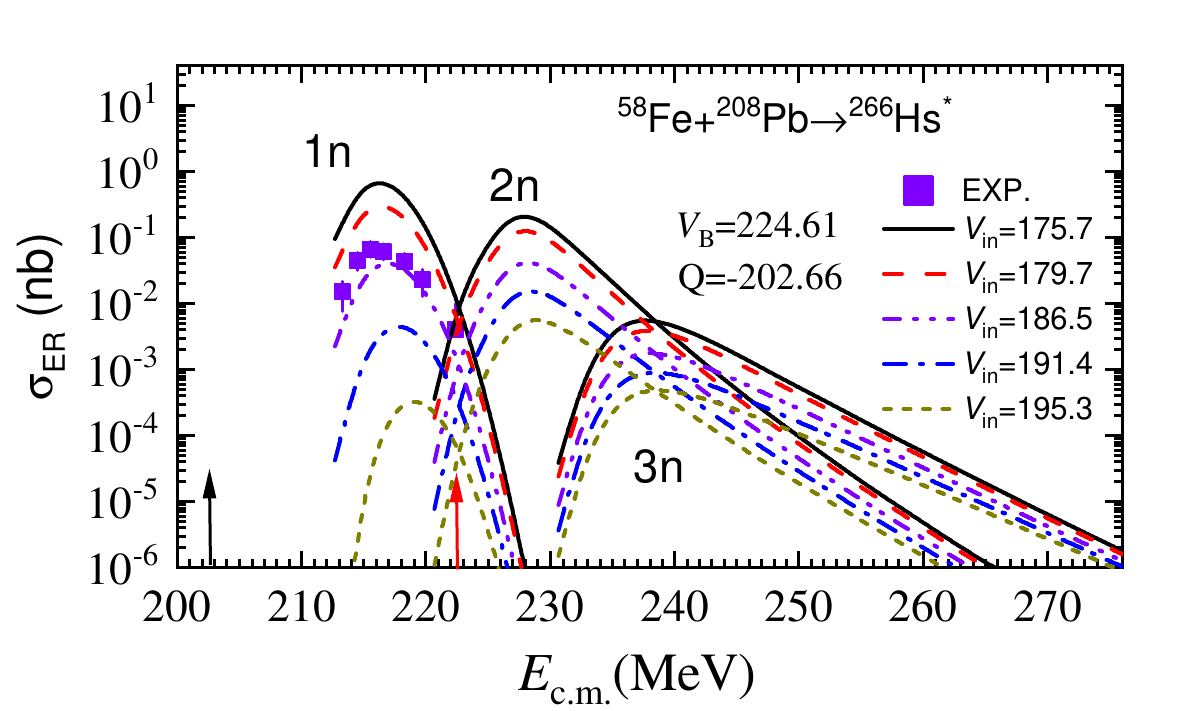}
\caption{\label{fig2} The calculated evaporation residual cross section (ERCS) of $^{58}$Fe+$^{208}$Pb$\rightarrow$ $^{266}$Hs$^*$ are shown with potential energy at the contact point, compared with experimental data. 
The solid purple squares indicate the experimental data. The black solid, red dash, purple dash-dot, blue dash-dot-dot, and brown short-dash lines stand for the ERCS with $V_{\rm in}$=175.77 MeV, $V_{\rm in}$=179.67 MeV, $V_{\rm in}$=186.51 MeV, $V_{\rm in}$=191.39 MeV, $V_{\rm in}$=195.3 MeV, respectively. The solid black and red arrows stand for Q value and Bass barrier.}
\end{figure}

To investigate the impact of contact potential energy on the excitation functions of Evaporation Residue cross sections (ERCS), we conduct calculations for the reaction $^{58}$Fe + $^{208}$Pb $\rightarrow$ $^{266}$Hs$^*$ using varying $V_{\rm in}$ values as depicted in Fig. \ref{fig2}. These calculations were performed within the DNS model to identify the most effective $V_{\rm in}$.
Fig. \ref{fig2} displays the calculated excitation functions for the 1n-, 2n-, and 3n-channel ERCS, represented by lines of the same style. Distinct line styles correspond to different $V_{\rm in}$ values. The solid black and red arrows indicate the Q value and the Bass barrier, respectively. The Bass potential is calculated in detail as per \cite{PhysRevLett.39.265}. The solid purple squares with error bars correspond to experimental data, which are observed to lie below the Bass barrier, suggesting that fusion-evaporation reactions predominantly occur sub-barrier. The 1n-channel ERCS functions exhibit greater shifts than the 2n- or 3n-channels, indicating that sub-barrier fusion is sensitive to variations in the contact potential energy.
The optimal contact potential energy, identified from Fig. \ref{fig2}, is $V_{\rm in} = 186.5$ MeV, which is denoted by the purple dash-dot-dot line.

\begin{table}
\renewcommand{\arraystretch}{1.2}
\centering
\setlength{\tabcolsep}{6pt}
  \caption{\label{tab1} Available combinations of projectile and target materials could be used to synthesize the new SHN via cold-fusion reactions. The bold red compound could be used to synthesize new SHN by 1n and 2n evaporation channels. Superscript asterisks on compound nuclei indicate the presence of experimental data.}
  \begin{tabular}{c|ccccc}
\toprule
    P $\backslash$ T  & $^{204}$Pb & $^{206}$Pb & $^{207}$Pb & $^{208}$Pb & $^{209}$Bi  \\
    \hline
    $^{46}$Ti & \textbf{\textcolor{red}{$^{250}$Rf}} & \textbf{\textcolor{red}{$^{252}$Rf}} & \textbf{\textcolor{red}{$^{253}$Rf}} & \textbf{\textcolor{red}{$^{254}$Rf}} & \textbf{\textcolor{red}{$^{255}$Db}} \\
    $^{47}$Ti & \textbf{\textcolor{red}{$^{251}$Rf}} & \textbf{\textcolor{red}{$^{253}$Rf}} & \textbf{\textcolor{red}{$^{254}$Rf}}& $^{255}$Rf & \textbf{\textcolor{red}{$^{256}$Db}} \\
    $^{48}$Ti & \textbf{\textcolor{red}{$^{252}$Rf}} & \textbf{\textcolor{red}{$^{254}$Rf}} & $^{255}$Rf  & $^{256}$Rf & $^{257}$Db \\
    $^{49}$Ti & \textbf{\textcolor{red}{$^{253}$Rf}}& $^{255}$Rf & $^{256}$Rf  & $^{257}$Rf & $^{258}$Db \\
    $^{50}$Ti & \textbf{\textcolor{red}{$^{254}$Rf}}& $^{256}$Rf & $^{257}$Rf  & $^{258}$Rf$^*$ & $^{259}$Db$^*$ \\
    \hline
    $^{50}$V & \textbf{\textcolor{red}{$^{254}$Db}} & \textbf{\textcolor{red}{$^{256}$Db}} & $^{257}$Db  & $^{258}$Db & \textbf{\textcolor{red}{$^{259}$Sg}} \\
    $^{51}$V & \textbf{\textcolor{red}{$^{255}$Db}} & $^{257}$Db & $^{258}$Db  & $^{259}$Db$^*$ & $^{260}$Sg \\
    \hline
    $^{50}$Cr & \textbf{\textcolor{red}{$^{254}$Sg}} & \textbf{\textcolor{red}{$^{256}$Sg}} &  \textbf{\textcolor{red}{$^{257}$Sg}} & \textbf{\textcolor{red}{$^{258}$Sg}} & \textbf{\textcolor{red}{$^{259}$Bh}} \\
    $^{52}$Cr & \textbf{\textcolor{red}{$^{256}$Sg}} & \textbf{\textcolor{red}{$^{258}$Sg}} &  \textbf{\textcolor{red}{$^{259}$Sg}} & $^{260}$Sg$^*$ & $^{261}$Bh \\
    $^{53}$Cr & \textbf{\textcolor{red}{$^{257}$Sg}} & \textbf{\textcolor{red}{$^{259}$Sg}} & $^{260}$Sg  & $^{261}$Sg & $^{262}$Bh\\
    $^{54}$Cr & \textbf{\textcolor{red}{$^{258}$Sg}} & $^{260}$Sg$^*$ & $^{261}$Sg$^*$  & $^{262}$Sg$^*$ & $^{263}$Bh$^*$ \\
    \hline
    $^{55}$Mn & \textbf{\textcolor{red}{$^{259}$Bh}} & \textbf{\textcolor{red}{$^{261}$Bh}} & $^{262}$Bh  & $^{263}$Bh$^*$ & \textbf{\textcolor{red}{$^{264}$Hs}} \\
    \hline
    $^{54}$Fe & \textbf{\textcolor{red}{$^{258}$Hs}} & \textbf{\textcolor{red}{$^{260}$Hs}} & \textbf{\textcolor{red}{$^{261}$Hs}}  & \textbf{\textcolor{red}{$^{262}$Hs}} & \textbf{\textcolor{red}{$^{263}$Mt}} \\
    $^{56}$Fe & \textbf{\textcolor{red}{$^{260}$Hs}} & \textbf{\textcolor{red}{$^{262}$Hs}} &\textbf{\textcolor{red}{$^{263}$Hs}} &\textbf{\textcolor{red}{$^{264}$Hs}}& \textbf{\textcolor{red}{$^{265}$Mt}}\\
    $^{57}$Fe &  \textbf{\textcolor{red}{$^{261}$Hs}}& \textbf{\textcolor{red}{$^{263}$Hs}}&\textbf{\textcolor{red}{$^{264}$Hs}}& $^{265}$Hs &\textbf{\textcolor{red}{$^{266}$Mt}} \\
    $^{58}$Fe & \textbf{\textcolor{red}{$^{262}$Hs}} &\textbf{\textcolor{red}{$^{264}$Hs}} & $^{265}$Hs  & $^{266}$Hs$^*$ &$^{267}$Mt$^*$ \\
    \hline
    $^{59}$Co & \textbf{\textcolor{red}{$^{263}$Mt}} & \textbf{\textcolor{red}{$^{265}$Mt}} &\textbf{\textcolor{red}{$^{266}$Mt}} & $^{267}$Mt$^*$ &\textbf{\textcolor{red}{$^{268}$Ds}}\\
    \hline
    $^{58}$Ni & \textbf{\textcolor{red}{$^{262}$Ds}} &  \textbf{\textcolor{red}{$^{264}$Ds}}& \textbf{\textcolor{red}{$^{265}$Ds}}  & \textbf{\textcolor{red}{$^{266}$Ds}} & \textbf{\textcolor{red}{$^{267}$Rg}} \\
    $^{60}$Ni & \textbf{\textcolor{red}{$^{264}$Ds}} & \textbf{\textcolor{red}{$^{266}$Ds}} &\textbf{\textcolor{red}{$^{267}$Ds}} & \textbf{\textcolor{red}{$^{268}$Ds}} &  \textbf{\textcolor{red}{$^{269}$Rg}}\\
    $^{61}$Ni & \textbf{\textcolor{red}{$^{265}$Ds}} & \textbf{\textcolor{red}{$^{267}$Ds}} &\textbf{\textcolor{red}{$^{268}$Ds}} & $^{269}$Ds & \textbf{\textcolor{red}{$^{270}$Rg}}\\
    $^{62}$Ni & \textbf{\textcolor{red}{$^{266}$Ds}} &\textbf{\textcolor{red}{$^{268}$Ds}} & $^{269}$Ds  & $^{270}$Ds$^*$ & \textbf{\textcolor{red}{$^{271}$Rg}} \\
    $^{64}$Ni &\textbf{\textcolor{red}{$^{268}$Ds}} & $^{270}$Ds & $^{271}$Ds$^*$  & $^{272}$Ds&$^{273}$Rg$^*$\\
    \hline
    $^{63}$Cu & \textbf{\textcolor{red}{$^{267}$Rg}} & \textbf{\textcolor{red}{$^{269}$Rg}} & \textbf{\textcolor{red}{$^{270}$Rg}} & \textbf{\textcolor{red}{$^{271}$Rg}}& \textbf{\textcolor{red}{$^{272}$Cn}} \\
    $^{65}$Cu & \textbf{\textcolor{red}{$^{269}$Rg}} & \textbf{\textcolor{red}{$^{271}$Rg}} & \textbf{\textcolor{red}{$^{272}$Rg}} & \textbf{\textcolor{red}{$^{273}$Rg}}& \textbf{\textcolor{red}{$^{274}$Cn}} \\
    \hline
    $^{64}$Zn & \textbf{\textcolor{red}{$^{278}$Cn}} &\textbf{\textcolor{red}{$^{270}$Cn}} & \textbf{\textcolor{red}{$^{271}$Cn}} &\textbf{\textcolor{red}{$^{272}$Cn}}& \textbf{\textcolor{red}{$^{273}$Nh}} \\
    $^{66}$Zn & \textbf{\textcolor{red}{$^{270}$Cn}}&\textbf{\textcolor{red}{$^{272}$Cn}} &\textbf{\textcolor{red}{$^{273}$Cn}}&\textbf{\textcolor{red}{$^{274}$Cn}}& \textbf{\textcolor{red}{$^{275}$Nh}} \\
    $^{67}$Zn &\textbf{\textcolor{red}{$^{271}$Cn}}&\textbf{\textcolor{red}{$^{273}$Cn}}&\textbf{\textcolor{red}{$^{274}$Cn}}&\textbf{\textcolor{red}{$^{275}$Cn}}& \textbf{\textcolor{red}{$^{276}$Nh}} \\
    $^{68}$Zn &\textbf{\textcolor{red}{$^{272}$Cn}}&\textbf{\textcolor{red}{$^{274}$Cn}} & \textbf{\textcolor{red}{$^{275}$Cn}} &\textbf{\textcolor{red}{$^{276}$Cn}}& \textbf{\textcolor{red}{$^{277}$Nh}} \\
    $^{70}$Zn &\textbf{\textcolor{red}{$^{274}$Cn}} & \textbf{\textcolor{red}{$^{276}$Cn}} & \textbf{\textcolor{red}{$^{277}$Cn}}& $^{278}$Cn$^*$ &$^{279}$Nh$^*$\\
    \bottomrule
    \end{tabular}
    \end{table}

\begin{figure*}[htb]
\includegraphics[width=.88\linewidth]{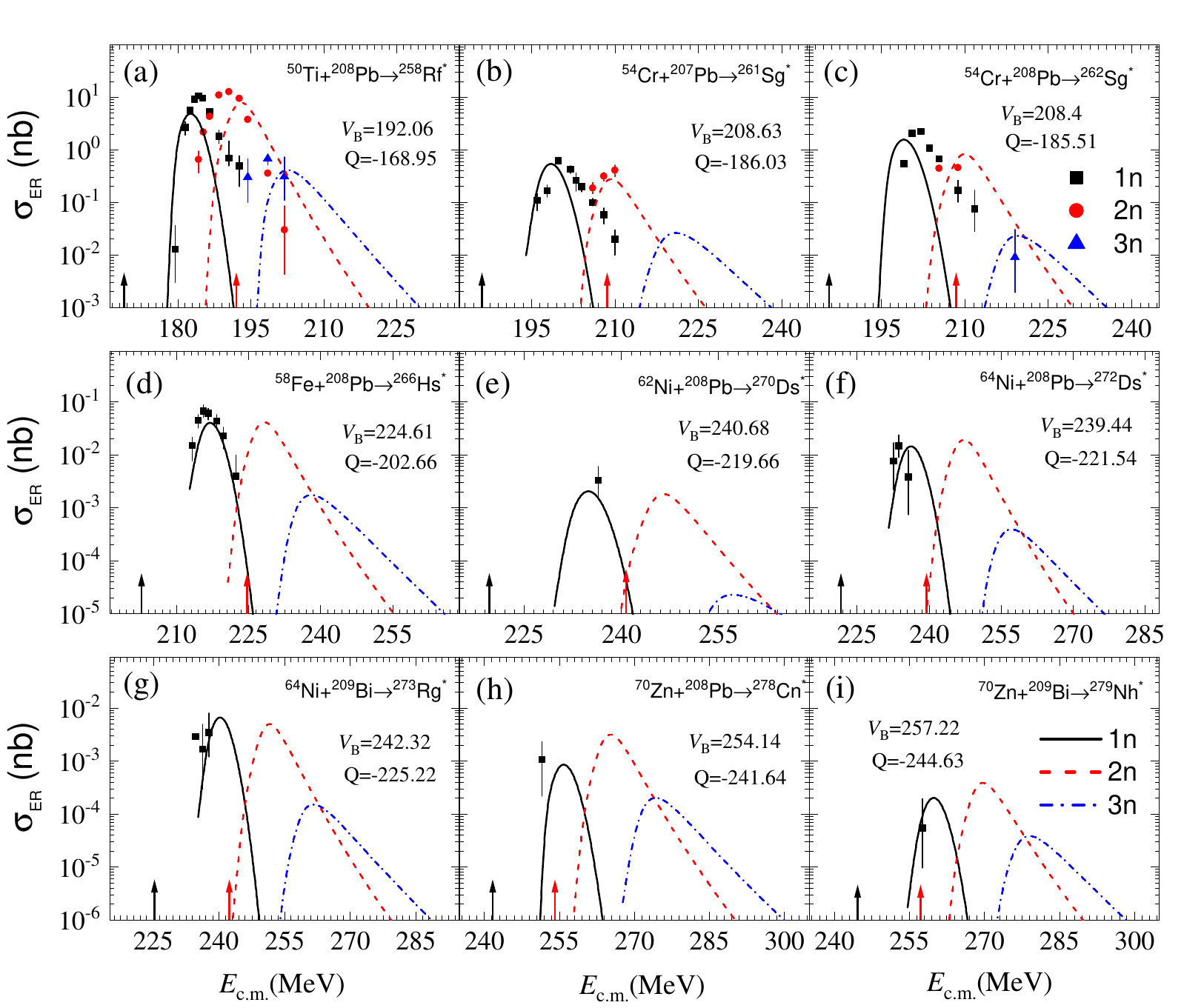}
\caption{\label{fig3}
Panels (a) to (i) depict the calculated excitation functions for Evaporation Residue cross sections (ERCS) resulting from cold-fusion reactions of projectile nuclei $^{50}$Ti, $^{54}$Cr, $^{58}$Fe, $^{62,64}$Ni, and $^{70}$Zn with target nuclei $^{207,208}$Pb and $^{209}$Bi, juxtaposed with experimental data indicated by error bars \cite{hessberger2001decay,mitsuoka2007barrier,RevModPhys.72.733,hofmann2000discovery,hofmann1995new,feng2007formation}.
The calculated ERCS for the 1n-, 2n-, and 3n-channels are represented by black solid, red dashed, and blue dash-dot lines, respectively. Correspondingly, experimental data for the 1n-, 2n-, and 3n-channels are marked by black solid squares, red circles, and blue up-triangles, respectively. The black and red arrows in the figure signify the Q value and Bass potential, respectively.
}
\end{figure*}

\begin{figure*}[htb]
\includegraphics[width=0.999\linewidth]{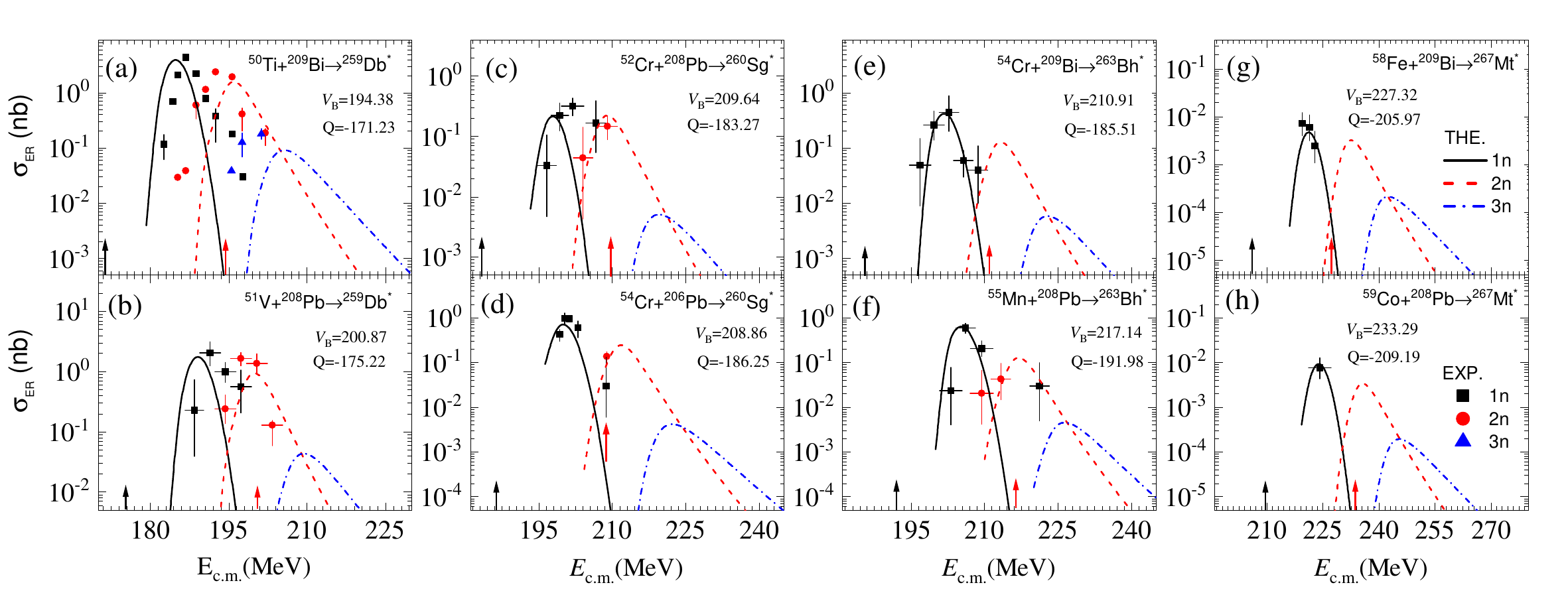}
\caption{\label{fig4} 
The panels are analogous to those in Figure \ref{fig4}, but pertain to the cold-fusion reactions involving projectile nuclei $^{50}$Ti, $^{51}$V, $^{52,54}$Cr, $^{55}$Mn, $^{58}$Fe, and $^{59}$Co incident on target nuclei $^{206,208}$Pb and $^{209}$Bi. The experimental data for these reactions are cited from \cite{PhysRevC.78.034604,hessberger2001decay,PhysRevC.78.024606,PhysRevC.79.027605}. Each column corresponds to the same compound nuclei, namely $^{259}$Db, $^{260}$Sg, $^{263}$Bh, and $^{267}$Mt, arranged from left to right.
}
\end{figure*}

\begin{figure*}[htb]
\includegraphics[width=0.85\linewidth]{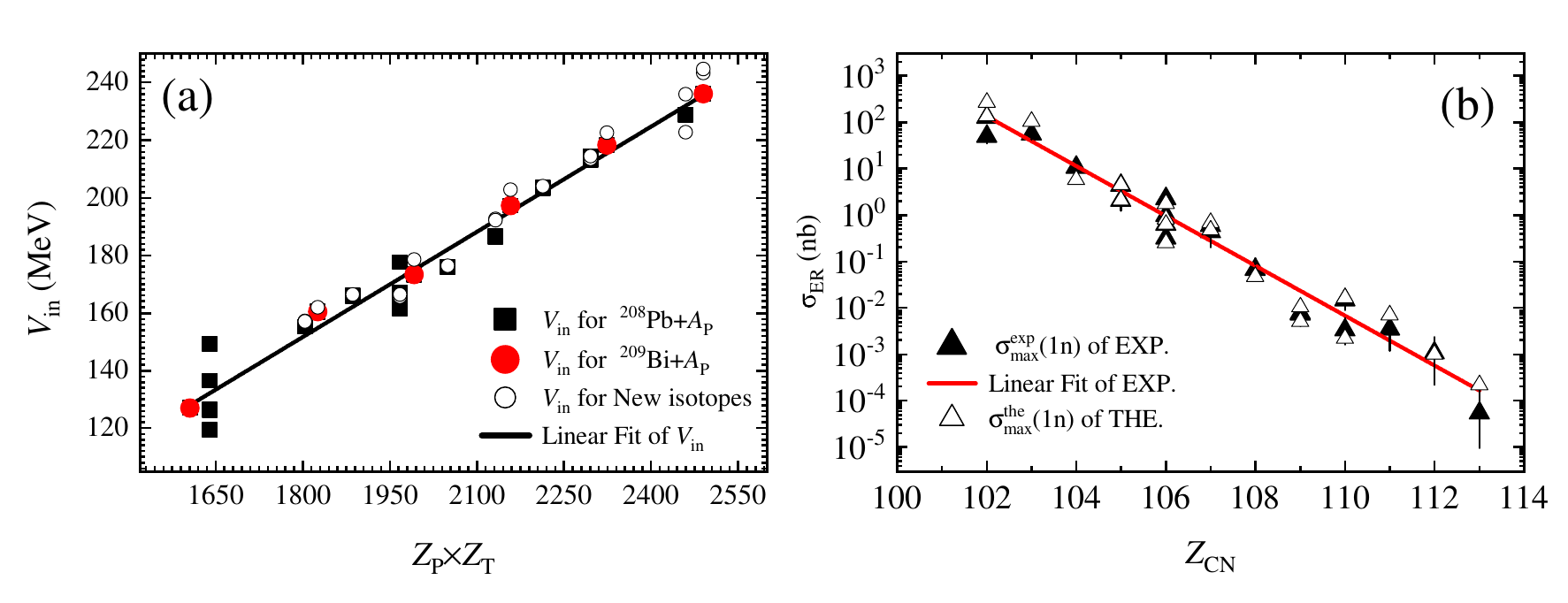}
\caption{\label{fig5} 
Panel (a) presents the shift in interaction potential energies \( V_{\rm in} \) at the contact point, as the key input quantities in the DNS model, across various reaction systems. These shifts are plotted against the product of atomic numbers \( Z_{\rm P} \times Z_{\rm T} \) and are utilized in the calculations for SHN with atomic numbers \( Z \) ranging from 102 to 113. The data are aligned with cold-fusion experimental observations, with a linear fit represented by the solid black line. Solid black squares and red circles correspond to reactions targeting \( ^{208} \)Pb and \( ^{209} \)Bi, respectively.
Panel (b) illustrates the maximum cross sections calculated for the 1n channel in the synthesis of SHN with \( Z \) from 102 to 113. These calculations are juxtaposed with experimental data, indicated by black open and solid triangles, each accompanied by an error bar. The red line delineates the linear fit to the experimental maximum cross sections \( \sigma^{\rm exp}_{\rm max} \).
}
\end{figure*}

Figure \ref{fig3} illustrates the calculated excitation functions for ERCS in cold-fusion reactions, as predicted by the DNS model, in comparison with experimental data. The black solid, red dashed, and blue dash-dot lines represent the calculated ERCS for the 1n-, 2n-, and 3n-channels, respectively. Correspondingly, the black solid squares, red circles, and blue up-triangles with error bars denote the experimental ERCS data for the 1n-, 2n-, and 3n-channels. The Q value and Bass barrier are indicated by the solid black and red arrows, respectively.
Our calculations are found to be in good agreement with the experimental data. The figure details the excitation functions for a series of nuclear reactions, each proceeding through 1n to 3n evaporation channels, with the associated Q-values, contact potentials ($V_{\rm in}$), minimum potentials ($V_{\rm S}$), and Bass barriers ($V_{\rm B}$) provided for each reaction.
It is observed that $V_{\rm in}$ is less than $V_{\rm S}$ in Panels (a)-(f), suggesting that nucleon transfer occurs prior to surpassing the Coulomb barrier, consistent with previous research \cite{Cook2023}.
From Fig. \ref{fig4} (a) to (c), the experimental data are available for the 1n- and 2n-channels ERCS, while the remaining panels present the 1n channel only. The 1n channel consistently exhibits larger ERCS compared to other neutron channels. The maxima of ERCS shift to lower energies with increasing mass asymmetry between the projectile and target, attributed to two factors: the larger inner fusion barrier associated with greater mass asymmetry, and the smaller fission barrier of heavier compound nuclei formed with a fixed target (Bi, Pb), leading to a decreased survival probability. The 3n ERCS is significantly smaller than the 1n or 2n channels by 1-2 orders of magnitude, due to the exponential decrease in fission barriers of compound nuclei with increasing excitation energy.
The majority of these reactions exhibit their largest ERCS below the Bass potential $V_{\rm B}$. The applied contact potential energies $V_{\rm in}$ for these reactions are compiled and displayed in Fig. \ref{fig5}.

Figure \ref{fig4} presents the calculated excitation functions for the Evaporation Residue cross sections (ERCS) of eight distinct nuclear reactions, categorized into four columns, and compared against available experimental data. The reactions are as follows:
1) $^{51}$V + $^{208}$Pb and $^{50}$Ti + $^{209}$Bi, both yielding $^{259}$Db.
2) $^{52}$Cr + $^{208}$Pb and $^{54}$Cr + $^{206}$Pb, both resulting in $^{260}$Sg.
3) $^{54}$Cr + $^{209}$Bi and $^{55}$Mn + $^{208}$Pb, leading to $^{263}$Bh.
4) $^{58}$Fe + $^{209}$Bi and $^{59}$Co + $^{208}$Pb, culminating in $^{267}$Mt.
The first column compares reactions producing the same compound nucleus $^{259}$Db, allowing for the examination of projectile-target effects on ERCS without the complication of de-excitation processes. The reaction $^{51}$V + $^{208}$Pb exhibits a maximum ERCS of $\sigma^{(\rm 4b)}_{\rm max}$(1n) = 2.05 nb, whereas $^{50}$Ti + $^{209}$Bi shows a larger maximum of $\sigma^{(\rm 4a)}_{\rm max}$(1n) = 4.36 nb. The difference in Bass potential energies, $V_{\rm B}$ = 194.38 MeV and $V_{\rm B}$ = 200.87 MeV, respectively, results in a ratio of $\sigma^{(\rm 4b)}_{\rm max}$(1n)/$\sigma^{(\rm 4a)}_{\rm max}$(1n) = 0.47, indicative of a significant impact of projectile and target composition on ERCS.
The second column examines ERCS for the reactions $^{52}$Cr + $^{208}$Pb and $^{54}$Cr + $^{206}$Pb, which form the same compound nucleus $^{260}$Sg. The reaction with $^{52}$Cr + $^{208}$Pb has a maximum ERCS of $\sigma^{(\rm 4c)}_{\rm max}$(1n) = 0.32 nb, while the reaction with $^{54}$Cr + $^{206}$Pb has a maximum of $\sigma^{(\rm 4d)}_{\rm max}$(1n) = 0.98 nb. The slight discrepancy in Bass potential energies, $V_{\rm B}$ = 209.64 MeV and $V_{\rm B}$ = 208.86 MeV, leads to a ratio of $\sigma^{(\rm 4c)}_{\rm max}$(1n)/$\sigma^{(\rm 4d)}_{\rm max}$(1n) = 0.32, underscoring the sensitivity of ERCS to the neutron content of the projectile and target nuclei.

Further analysis reveals that neutron-rich projectile nuclei, when combined with a fixed target ($^{208}$Pb), tend to achieve larger ERCS. This may be attributed to the reduced Coulomb barrier in neutron-rich systems and a higher survival probability for the compound SHN through neutron evaporation. Additionally, the comparison between reactions with varying target neutron content, while maintaining a fixed projectile nucleus ($^{54}$Cr), shows that larger neutron-rich target nuclei can enhance ERCS. However, the odd-even effect may counteract or even surpass the benefits of neutron richness.
The findings suggest that both projectile and target neutron content play a crucial role in determining ERCS, with neutron-rich combinations favoring larger cross sections, likely due to the interplay between reduced Coulomb barriers and increased stability through neutron evaporation.

Figure \ref{fig5}(a) depicts the trend of contact potential energies, denoted as $V_{\rm in}$, in relation to the product of the charge numbers of the projectile nucleus ($Z_{\rm P}$) and the target nucleus ($Z_{\rm T}$). The black solid squares and red solid circles correspond to $V_{\rm in}$ values for reactions based on $^{208}$Pb and $^{209}$Bi targets, respectively. A linear fit to $V_{\rm in}$ is represented by the solid black line, demonstrating a linear increase of $V_{\rm in}$ with the product \( Z_{\rm P} \times Z_{\rm T} \). Panel (b) presents the maximum ERCS values, $\sigma_{\rm max}$(1n), with solid up-triangles and open up-triangles signifying experimental data ($\sigma^{\rm exp}_{\rm max}$(1n)) and calculated results ($\sigma^{\rm the}_{\rm max}$(1n)), respectively. The red solid line represents a linear fit to the experimental maximum ERCS values, revealing a strong correlation between our calculations and experimental findings. The trend indicates a linear decrease in the maximum ERCS with the charge number of the compound SHN.
Employing this established relationship, we can extrapolate the synthesis cross sections for unknown superheavy isotopes formed via cold-fusion reactions. The open black circles in the figure denote the $V_{\rm in}$ values for reactions predicted to yield new SHN, providing a basis for estimating their ERCS.
    
\begin{figure*}[htb]
\includegraphics[width=.999\linewidth]{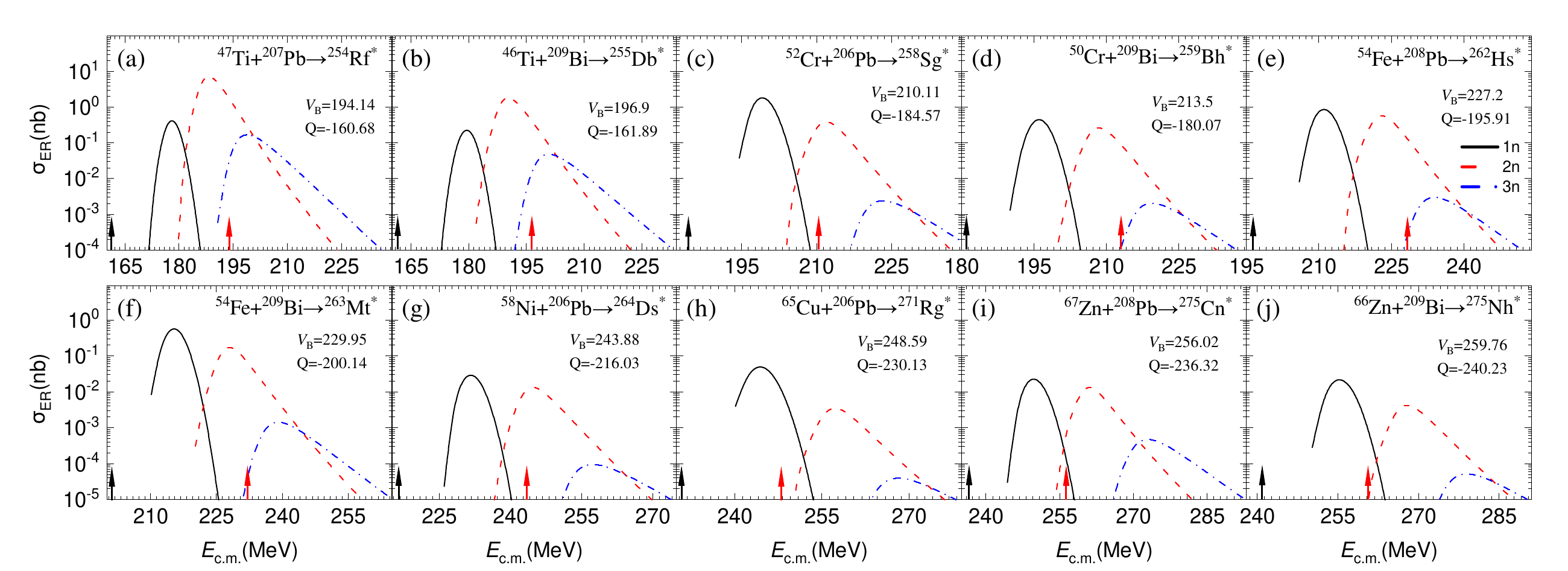}
\caption{\label{fig6}(Color online)
The predicted excitation energies of ERCS for the cold-fusion reactions of SHN with atomic numbers 104 to 113 are presented in panels (a) through (j), consistent with the arrangement in Figure \ref{fig4}.
}
\end{figure*}

Table \ref{tab1} enumerates all 145 combinations of projectile-target nuclei derived from natural elements. These combinations are poised to probe previously unexplored regions, as indicated by the red, bolded entries. Utilizing our model, we have conducted calculations for each of these 145 combinations. Notably, 10 of these combinations have been identified as optimal, selected for their larger synthesis cross sections, which are depicted in Fig. \ref{fig6}. 

Figure \ref{fig6} presents the calculated excitation functions for the ERCS of superheavy isotopes with atomic numbers ranging from 104 to 113, which are synthesized through optimal cold-fusion reactions. These optimal combinations offer a promising avenue for the production of new SHN within the specified atomic number range. These calculations are grounded in the DNS model, leveraging the contact potential energies ($V_{\rm in}$) as determined from the data presented in Figure \ref{fig5}. The selected projectile-target combinations consist of stable, naturally abundant nuclei, including $^{46,47}$Ti, $^{51}$V, $^{50,52}$Cr, $^{54}$Fe, $^{58}$Ni, $^{65}$Cu and $^{66,67}$Zn as projectile materials, with $^{204,206,207,208}$Pb and $^{209}$Bi serving as target materials, shown in panels (a)-(j). 
The black solid, red dashed, and blue dash-dot lines correspond to the 1n-, 2n-, and 3n-channels of the ERCS, respectively.
The black arrows and red arrows stand for the Q-value and Bass potential energies.
The contact potential $V_{\rm in}$ applied in calculations for specific reactions are listed in Fig. \ref{fig5}, marked by open black circles.
The prediction of the maximum ERCS for unknown SHN, which are listed in Table \ref{tab2}.
Our findings indicate that a substantial number of previously unknown SHN can be synthesized with a high probability through cold fusion reactions, utilizing stable projectile-target nucleus combinations. Consequently, cold fusion reactions retain significant potential for the synthesis of SHN within the atomic number range of 104 to 113.

\section{Conclusion}\label{sec4}

To test whether the cold fusion reaction method still holds the potential for synthesizing SHN, we systematically calculated reactions for SHN with proton numbers in the range of 104-113, within the framework of the DNS model, based on 145 projectile-target combinations using naturally occurring nuclei as projectile-target materials, which listed in Table \ref{tab1}.
Cold fusion reactions for synthesizing SHN are predominantly characterized by maximum cross sections in the 1n to 2n evaporation channels. This is primarily attributed to the product of the survival probability and the fusion probability peaking near the Coulomb barrier, with the sum of the Coulomb barrier and the Q-value falling within 30 MeV. We observed that the maximum cross section for cold fusion occurs below the Bass barrier, which can be interpreted in two ways: either the Bass barrier is not suitable for cold fusion reactions, or significant deformation in the cold fusion system leads to a reduced effective Coulomb barrier. Before predicting the synthesis of SHN, we systematically calculated reaction systems with existing experimental data and found good agreement with these data.
Reaction systems with similar mass for the synthesized compound nucleus can be used to study the entrance channel effects. For reactions systems with the same projectile-target element synthesizing the same compound nucleus, we found that more neutron-rich projectiles are slightly more favorable for fusion, although the effect is minor. For a fixed target nucleus (Pb, Bi), the evaporation residue cross section shows a linear decrease with the increasing proton number of the projectile, as depicted in Fig. \ref{fig5} (b). The main reasons for this trend are twofold: first, the fusion probability drops rapidly with increasing asymmetry between the projectile and target; second, SHN with a larger atomic number have a lower fission barrier, leading to a decreased survival probability during the de-excitation process. 
Since the potential energy at the contact point is dynamically changing before nucleon transfer between the projectile and target nuclei, within the DNS model, we systematically compared with experimental data and extracted the contact potential energies $V_{\rm in}$ at the contact of the projectile-target nuclei, as shown in Fig. \ref{fig5}. We observed a linear increasing trend of $V_{\rm in}$ with the product of the proton numbers of the projectile-target system, which was determined through linear fitting. The Vin for the reaction system depends on the isotopes of the projectile-target nuclei, with a slightly lower $V_{\rm in}$ observed for more neutron-rich systems. For some reaction systems, we found $V_{\rm in}$ values to be less than the potential energy $V_{\rm S}$, which may be explained by nucleon transfer occurring before capture, leading to an increased asymmetry in the projectile nucleus. Based on the DNS model and the $V_{\rm in}$ distribution discovered through comparison with experimental data, for the nuclear region with Z=104-113, we systematically calculated 145 kinds of projectile-target combinations and found that SHN have considerable synthesis cross sections. The excitation functions are listed in Fig. \ref{fig6}, and the optimal projectile-target combinations, the best collision energies, and the maximum synthesis cross section information are presented in Table \ref{tab2}. In summary, we believe that cold fusion reactions hold significant potential for synthesizing proton-rich new SHN near the drip line in the Z=104-113 region, with clear advantages over other reaction mechanisms.

\begin{table*}
\renewcommand{\arraystretch}{1.2}
\centering
\setlength{\tabcolsep}{2pt}
  \caption{\label{tab2}New superheavy isotopes within the atomic number range of 104 to 113 have been projected. The optimal reactions, Bass potentials, collision energies (excitation energies), and synthesis cross sections are listed sequentially.}
  \begin{tabular}{cccc|cccc}
    \toprule
    Reactions  &$V_{\rm B}$(MeV) &$E_{\rm c.m.}$(MeV) & $\sigma_{\rm max}$(nb) & Reactions  &$V_{\rm B}$(MeV) &$E_{\rm c.m.}$(MeV) & $\sigma_{\rm max}$(nb) \\
    \hline
    $^{46}$Ti($^{204}$Pb, 2n)$^{248}$Rf & 195.43 & 192.72 (30) & 1.227    &  $^{58}$Ni($^{206}$Pb, 1n)$^{263}$Ds & 243.88 &232.03 (16)  & 0.033\\    
    $^{47}$Ti($^{204}$Pb, 2n)$^{249}$Rf & 194.79 &190.54 (26)  & 1.041    &  $^{58}$Ni($^{207}$Pb, 1n)$^{264}$Ds & 243.61 &228.89 (14)  & $8.1\times 10^{-3}$\\
    $^{46}$Ti($^{206}$Pb, 2n)$^{250}$Rf & 195    &189.08 (28)  & 5.124    &  $^{58}$Ni($^{208}$Pb, 1n)$^{265}$Ds & 243.34 &229.42 (16)  & $8.6\times 10^{-3}$\\
    $^{48}$Ti($^{204}$Pb, 1n)$^{251}$Rf & 194.15 &183.46 (16)  & 4.836    &  $^{60}$Ni($^{207}$Pb, 1n)$^{266}$Ds & 242.26 &232.89 (14)  & $6.2\times 10^{-3}$\\
    $^{47}$Ti($^{207}$Pb, 2n)$^{252}$Rf & 194.14 &188.68 (28)  & 7.697    &  $^{58}$Ni($^{209}$Bi, 2n)$^{265}$Rg & 246.28 &246.29 (28)  & $4.78\times 10^{-3}$\\
    $^{50}$V($^{204}$Pb, 2n)$^{252}$Db & 202.38 &200.73 (26)  & 0.186     &  $^{63}$Cu($^{204}$Pb, 1n)$^{266}$Rg & 250.44 &243.49 (14)  & 0.035\\
    $^{46}$Ti($^{209}$Bi, 2n)$^{253}$Db & 196.9 &189.89 (28)  & 2.046     &  $^{65}$Cu($^{204}$Pb, 2n)$^{267}$Rg & 249.14 &259.48 (28)  & $1.24\times 10^{-3}$\\
    $^{47}$Ti($^{209}$Bi, 2n)$^{254}$Db & 196.25 &189.85 (26)  & 1.655    &  $^{65}$Cu($^{204}$Pb, 1n)$^{268}$Rg & 245.48 &259.48 (14)  & 0.0249\\
    $^{50}$Cr($^{204}$Pb, 2n)$^{252}$Sg & 211.9 &211.06 (30)  & 0.1497    &  $^{65}$Cu($^{206}$Pb, 2n)$^{269}$Rg & 248.59 &256.13 (26)  & $3.7\times 10^{-3}$\\
    $^{50}$Cr($^{204}$Pb, 1n)$^{253}$Sg & 211.9 &197.06 (16)  & 0.433     &  $^{65}$Cu($^{206}$Pb, 1n)$^{270}$Rg & 248.59 &244.13 (14)  & 0.0574\\
    $^{50}$Cr($^{206}$Pb, 2n)$^{254}$Sg & 211.43 &206.95 (28)  & 0.29     &  $^{65}$Cu($^{207}$Pb, 1n)$^{271}$Rg & 248.32 &243.13 (14)  & 0.029\\
    $^{52}$Cr($^{204}$Pb, 1n)$^{255}$Sg & 210.58 &201.9 (16)  & 0.964     &  $^{64}$Zn($^{204}$Pb, 2n)$^{266}$Cn & 259.13 &264.87 (28)  & $2.3\times 10^{-4}$\\
    $^{53}$Cr($^{204}$Pb, 1n)$^{256}$Sg & 209.94 &199.82 (14)  & 0.709    &  $^{64}$Zn($^{204}$Pb, 1n)$^{267}$Cn & 259.13 &252.87 (16)  & $1\times 10^{-3}$\\
    $^{52}$Cr($^{206}$Pb, 1n)$^{257}$Sg & 210.11 &198.87 (14)  & 2.138    &  $^{66}$Zn($^{204}$Pb, 2n)$^{268}$Cn & 257.79 &267.72 (28)  & $3.5\times 10^{-4}$\\
    $^{50}$Cr($^{209}$Bi, 2n)$^{257}$Bh & 213.5 &208.07 (28)  & 0.3       &  $^{66}$Zn($^{204}$Pb, 1n)$^{269}$Cn & 257.79 &255.72 (16)  & $2.6\times 10^{-3}$\\
    $^{50}$Cr($^{209}$Bi, 1n)$^{258}$Bh & 213.5 &196.07 (16)  & 0.538     &  $^{66}$Zn($^{206}$Pb, 2n)$^{270}$Cn & 257.22 &266.03 (28)  & $1.4\times 10^{-3}$\\
    $^{55}$Mn($^{206}$Pb, 2n)$^{259}$Bh & 217.62 &219.16 (26)  & 0.096    &  $^{66}$Zn($^{207}$Pb, 2n)$^{271}$Cn & 256.94 &263.01 (26)  & $4.3\times 10^{-3}$\\
    $^{54}$Fe($^{208}$Pb, 2n)$^{260}$Hs & 227.25 &221.91 (26)  & 0.615    &  $^{66}$Zn($^{207}$Pb, 1n)$^{272}$Cn & 256.94 &251.01 (14)  & 0.0152\\
    $^{54}$Fe($^{208}$Pb, 1n)$^{261}$Hs & 227.25 &211.91 (16)  & 0.922    &  $^{67}$Zn($^{208}$Pb, 2n)$^{273}$Cn & 256.02 &260.32 (24)  & 0.0153\\
    $^{56}$Fe($^{207}$Pb, 1n)$^{262}$Hs & 226.13 &211.19 (14)  & 0.286    &  $^{67}$Zn($^{208}$Pb, 1n)$^{274}$Cn & 256.02 &250.32 (14)  & 0.0251\\
    $^{54}$Fe($^{209}$Bi, 2n)$^{261}$Mt & 229.95 &228.14 (28)  & 0.19     &  $^{68}$Zn($^{208}$Pb, 1n)$^{275}$Cn & 255.38 &252.82 (14)  & $8.8\times 10^{-3}$\\
    $^{54}$Fe($^{209}$Bi, 1n)$^{262}$Mt & 229.95 &216.14 (16)  & 0.6      &  $^{64}$Zn($^{209}$Bi, 2n)$^{271}$Nh & 261.11 &264.89 (28)  & $2.4\times 10^{-4}$\\
    $^{56}$Fe($^{209}$Bi, 2n)$^{263}$Mt & 228.06 &230.04 (26)  & 0.011    &  $^{64}$Zn($^{209}$Bi, 1n)$^{272}$Nh & 261.11 &252.89 (16)  & $9.3\times 10^{-4}$\\
    $^{56}$Fe($^{209}$Bi, 1n)$^{264}$Mt & 228.06 &218.04 (14)  & 0.018    &  $^{66}$Zn($^{209}$Bi, 2n)$^{273}$Nh & 259.76 &268.23 (28)  & $4.58\times 10^{-3}$\\
    $^{58}$Ni($^{204}$Pb,2n)$^{260}$Ds  & 244.42 &247.96(30)   & $1.1\times 10^{-3}$ &  $^{66}$Zn($^{209}$Bi, 1n)$^{274}$Nh & 259.76 &254.23 (14)  & 0.0226\\
    $^{58}$Ni($^{204}$Pb, 1n)$^{261}$Ds & 244.42 &233.96 (16)  & 0.0124   &  $^{67}$Zn($^{209}$Bi, 1n)$^{275}$Nh & 259.11 &254.43 (14)  & $5.18\times 10^{-3}$\\    
    $^{58}$Ni($^{206}$Pb, 2n)$^{262}$Ds & 243.88 &244.03 (28)  & 0.015    &  $^{68}$Zn($^{209}$Bi, 1n)$^{276}$Nh & 258.47 &256.83 (14)  & 0.01\\
    \bottomrule
    \end{tabular}
    \end{table*}

\end{CJK}


%

\end{document}